\documentclass[
  aps,
 prapplied,
  superscriptaddress,
  floatfix,
  twocolumn,
  reprint,
  secnumarabic,
  amssymb,
  amsmath]{revtex4-1}

\usepackage{dcolumn}
\usepackage{bm}
\usepackage{hyperref}
\usepackage[pdftex]{graphicx}
\usepackage{tabularx}
\usepackage{subfigure}
\usepackage{amssymb}
\usepackage{enumerate}
\usepackage{array}
\usepackage{siunitx}
\newlength{\gnupicwidth}
\usepackage{ifthen}
\usepackage{color}
\usepackage{ulem}             
\definecolor{rot}{rgb}{1,0,0}
\definecolor{blau}{rgb}{0,0,1}
\definecolor{orange}{rgb}{.5,.5,0}
\definecolor{dunkelgruen}{rgb}{.133,0.545,0.133}

\newif\ifcom
\newif\ifdel
\comtrue               %
\delfalse

\begin{document}

\title{Large spin Hall magnetoresistance in antiferromagnetic $\alpha$-Fe$_2$O$_3$/Pt heterostructures}

\author{Johanna~Fischer}
\altaffiliation[Present address: ]{Unit\'{e} Mixte de Physique, CNRS, Thales, Universit\'{e} Paris-Sud, Universit\'{e} Paris-Saclay, 91767 Palaiseau, France}
\affiliation{Walther-Mei{\ss}ner-Institut, Bayerische Akademie der Wissenschaften, 85748 Garching, Germany}
\affiliation{Physik-Department, Technische Universit\"{a}t M\"{u}nchen, 85748 Garching, Germany}
\author{Matthias~Althammer}
\affiliation{Walther-Mei{\ss}ner-Institut, Bayerische Akademie der Wissenschaften, 85748 Garching, Germany}
\affiliation{Physik-Department, Technische Universit\"{a}t M\"{u}nchen, 85748 Garching, Germany}
\author{Nynke~Vlietstra}
\affiliation{Walther-Mei{\ss}ner-Institut, Bayerische Akademie der Wissenschaften, 85748 Garching, Germany}
\affiliation{Physik-Department, Technische Universit\"{a}t M\"{u}nchen, 85748 Garching, Germany}
\author{Hans~Huebl}
\affiliation{Walther-Mei{\ss}ner-Institut, Bayerische Akademie der Wissenschaften, 85748 Garching, Germany}
\affiliation{Physik-Department, Technische Universit\"{a}t M\"{u}nchen, 85748 Garching, Germany}
\author{Sebastian~T.B.~Goennenwein}
\affiliation{Institut f\"{u}r Festk\"{o}rper- und Materialphysik, Technische Universit\"{a}t Dresden, 01062 Dresden, Germany}
\affiliation{Center for Transport and Devices of Emergent Materials, Technische Universit\"{a}t Dresden, 01062 Dresden, Germany}
\author{Rudolf~Gross}
\affiliation{Walther-Mei{\ss}ner-Institut, Bayerische Akademie der Wissenschaften, 85748 Garching, Germany}
\affiliation{Physik-Department, Technische Universit\"{a}t M\"{u}nchen, 85748 Garching, Germany}
\affiliation{Munich Center for Quantum Science and Technology (MCQST), 80799 Munich, Germany}
\author{Stephan~Gepr\"{a}gs}
\affiliation{Walther-Mei{\ss}ner-Institut, Bayerische Akademie der Wissenschaften, 85748 Garching, Germany}
\author{Matthias~Opel}
\email[]{matthias.opel@wmi.badw.de}
\affiliation{Walther-Mei{\ss}ner-Institut, Bayerische Akademie der Wissenschaften, 85748 Garching, Germany}

\date{\today}

\begin{abstract} 
  We investigate the spin Hall magnetoresistance (SMR) at room temperature in thin film heterostructures of antiferromagnetic, insulating, (0001)-oriented $\alpha$-Fe$_2$O$_3$ (hematite) and Pt.
  We measure their longitudinal and transverse resistivities while rotating an applied magnetic field of up to 17\,T in three orthogonal planes.
  For out-of-plane magnetotransport measurements, we find indications for a multidomain antiferromagnetic configuration whenever the field is aligned along the film normal.
  For in-plane field rotations, we clearly observe a sinusoidal resistivity oscillation characteristic for the SMR due to a coherent rotation of the N\'{e}el vector.
  The maximum SMR amplitude of $0.25\%$ is, surprisingly, twice as high as for prototypical ferrimagnetic Y$_3$Fe$_5$O$_{12}$/Pt heterostructures.
  The SMR effect saturates at much smaller magnetic fields than in comparable antiferromagnets, making the $\alpha$-Fe$_2$O$_3$/Pt system particularly interesting for room-temperature antiferromagnetic spintronic applications.
\end{abstract}

\maketitle


\section{Introduction}

Despite lacking a net macroscopic magnetization, antiferromagnetic (AF) materials have moved into the focus of spintronics research \cite{Marti:2014, Wadley:2016, Jungwirth:2016, Baltz:2018}. Although L. N\'{e}el stated about 50~years ago that antiferromagnets ``are extremely interesting from the theoretical viewpoint, but do not seem to have any applications'' \cite{Neel:1970} this class of materials brings along two important advantages compared to ferromagnets: (i) they enable a better scalability and a higher robustness against magnetic field perturbations \cite{Marti:2014, Wadley:2016, Jungwirth:2016} and (ii) they offer orders of magnitudes faster dynamics and thus switching times \cite{Satoh:2014, Olejnik:2018}. Accordingly, AF spintronics emerged rapidly and brought out important developments ranging from random access memory schemes \cite{Kosub:2015, Kosub:2017} and the discovery of the spin colossal magnetoresistance \cite{Qiu:2018} in magnetoelectric antiferromagnets via synthetic antiferromagnetic spintronic devices \cite{Gomez-Perez:2018, Duine:2018} to the demonstration of long-range magnon spin transport in intrinsic antiferromagnets \cite{Lebrun:2018}.
From an application perspective, both switching the AF state as well as reading out the AF sublattice magnetization orientations are important challenges. It is evident that the vanishing net moment and the very small stray fields in AF materials call for new magnetization control and read-out strategies.

Spin currents \cite{Althammer:2018a} were shown to interact with individual magnetic sublattices via spin transfer torques, also in antiferromagnets \cite{Ando:2008, Miron:2011, Jia:2011, Liu:2012}. A particular manifestation of spin torque physics is the dependence of the resistivity of a metallic thin film with large spin-orbit coupling on the direction of the magnetization in an adjacent material with long range magnetic order, denoted as spin Hall magnetoresistance (SMR) effect \cite{Nakayama:2013, Vlietstra:2013, Althammer:2013, Chen:2013}. Following earlier results in all-metallic systems \cite{Kobs:2011}, the SMR was first established in oxide spintronics \cite{Coll:2019} for insulating, collinear ferrimagnetic Y$_3$Fe$_5$O$_{12}$/Pt bilayers \cite{Nakayama:2013, Vlietstra:2013, Althammer:2013}. Upon rotating the magnetization in the magnet/metal interface plane, the SMR appears as a sinusoidal oscillation of the Pt resistivity, characterized by a specific amplitude and a phase. In compensated ferrimagnetic YGd$_2$Fe$_4$InO$_{12}$/Pt heterostructures, the pronounced temperature dependence of the SMR phase demonstrated the sensitivity of the effect to the individual canted Fe$^{3+}$ \textit{sublattice} magnetizations \cite{Ganzhorn:2016}. Recently, the SMR effect was also identified in AF heterostructures. In spite of their zero net magnetization, the AF ordered magnetic sublattices contribute individually, resulting in a non-zero SMR. As the sublattice magnetizations are orthogonal to the applied magnetic field, a phase shift of $90^\circ$ was reported for the SMR in NiO/Pt \cite{Hoogeboom:2017, Fischer:2018, Baldrati:2018} and Cr$_2$O$_3$/Ta \cite{Ji:2018} as well as all-metallic PtMn/Pt \cite{DuttaGupta:2018} and PtMn/W \cite{DuttaGupta:2018} compared to that in the prototypical ferrimagnetic Y$_3$Fe$_5$O$_{12}$/Pt heterostructures. The SMR amplitude is still a matter of debate, since various extrinsic as well as intrinsic parameters play a crucial role \cite{Jia:2011, Meyer:2014, Althammer:2019}
and some authors report a non-zero amplitude even above the respective magnetic ordering temperatures \cite{Schlitz:2018, Velez:2019}.

In this Article, we substantially complement the SMR data available for AF insulators by investigating $\alpha$-Fe$_2$O$_3$/Pt. We find a surprisingly large SMR amplitude of $0.25\%$, much higher than in AF NiO/Pt \cite{Fischer:2018} and twice as large as in Y$_3$Fe$_5$O$_{12}$/Pt \cite{Althammer:2013}. This finding supports the picture that both AF sublattices contribute to the SMR at the interface, regardless of the material's net magnetization. The large SMR amplitude together with a moderate saturation field of $\sim\!3$\,T establishes $\alpha$-Fe$_2$O$_3$/Pt as a viable future SMR source and paves the way towards room temperature antiferromagnetic spintronic applications.

\section{Thin Film Deposition and Structural Characterization}

\begin{figure}
\includegraphics[width=1.0\columnwidth]{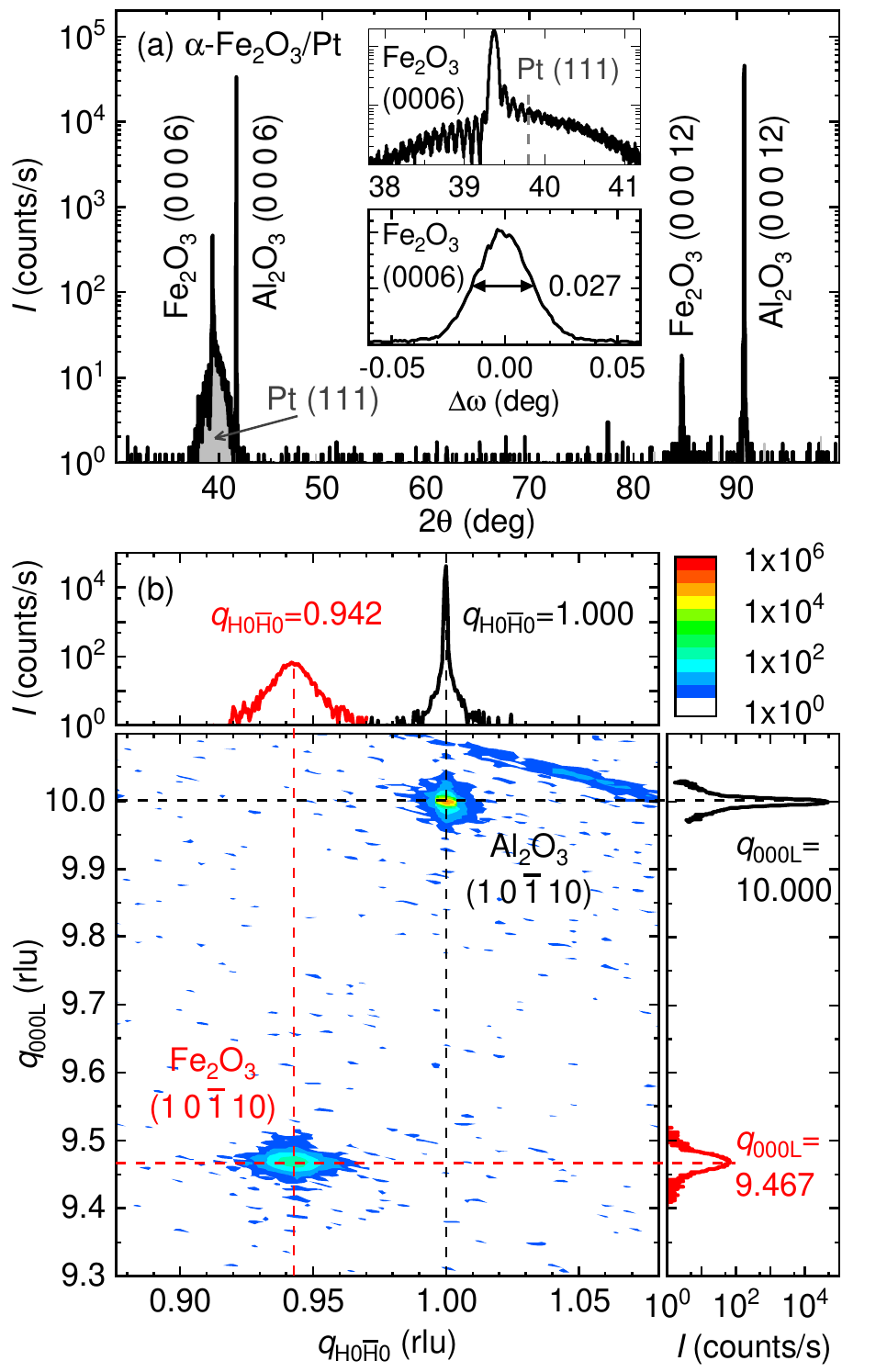}
    \caption{
    Structural properties of the investigated $\alpha$-Fe$_2$O$_3$/Pt heterostructure fabricated on a (0001)-oriented Al$_2$O$_3$ substrate.
    (a) 2$\theta$-$\omega$-scan along the [0001]-direction of Al$_2$O$_3$.
    The upper inset shows the scan on an enlarged scale together with the expected position of the Pt(111) reflection (vertical dashed line).
    The lower inset displays the rocking curve around the $\alpha$-Fe$_2$O$_3$(0006) reflection with a full width at half maximum of only $0.027^\circ$.
    (b) Reciprocal space mapping around the $(1\,0\,\overline{1}\,10)$ reflections. The reciprocal lattice units (rlu) are related to the Al$_2$O$_3$ substrate reflection.
 }
\label{fig:XRD}
\end{figure}

The electrical insulator $\alpha$-Fe$_2$O$_3$ (hematite) crystallizes in a rhombohedral structure and can be described in the hexagonal system with the lattice constants $a=0.5032$\,nm and $c=1.3748$\,nm \cite{Springer-Materials:Fe2O3}. In bulk, it exhibits a N\'{e}el temperature of $T_\mathrm{N}=953$\,K and undergoes a spin reorientation (``Morin'' transition) at $T_\mathrm{M} \approx 263$\,K \cite{Morin:1950}. For $T_\mathrm{M} < T < T_\mathrm{N}$ and in the absence of an external magnetic field, the $S=5/2$ spins of the Fe$^{3+}$ ions are ordered ferromagnetically in the (0001) planes. Along the crystallographic [0001] direction, these easy planes form a ``$+--+$'' sequence, resulting in a net AF order \cite{Shull:1951}. A finite anisotropic spin-spin (``Dzyaloshinskii-Moriya'') interaction \cite{Dzyaloshinskii:1958, Moriya:1960} leads to a small canting of the two AF sublattice magnetizations $\mathbf{M}_1$ and $\mathbf{M}_2$ with a canting angle of $0.13^\circ \pm 0.01^\circ$ \cite{Morrish:1994}. This results in a small net magnetization $\mathbf{M} = \mathbf{M}_1+\mathbf{M}_2$ in the (0001) plane. Similar to the situation in NiO \cite{Fischer:2018}, $\alpha$-Fe$_2$O$_3$ displays three AF domains rotated by $120^\circ$ with respect to each other \cite{Nathans:1964, Marmeggi:1977} and a domain population dependent on the direction and magnitude of the external magnetic field \cite{Marmeggi:1977}. The mono-domainization field $\mu_0H_\mathrm{MD}$ is reported to be above 600\,mT \cite{Marmeggi:1977}.

Since thin films are key for applications, we here study $\alpha$-Fe$_2$O$_3$/Pt bilayer heterostructures, fabricated on single crystalline, (0001)-oriented Al$_{2}$O$_{3}$ substrates. Using our pulsed laser deposition setup described in Ref.~\cite{Opel:2014}, we first deposit epitaxial $\alpha$-Fe$_2$O$_3$ thin films from a stoichiometric target with a laser fluence and a repetition rate of $2.5\,\mathrm{J/cm^2}$ and 2\,Hz, respectively, at a substrate temperature of $320^\circ$C in an oxygen atmosphere of $25\,\mu$bar. Without breaking the vacuum, the films are then covered $\textit{in-situ}$ by thin layers of Pt via electron beam evaporation. High-resolution X-ray diffractometry (HR-XRD) measurements reveal a high structural quality of the $\alpha$-Fe$_2$O$_3$/Pt heterostructures. The $2\theta$-$\omega$ scan (Fig.~\ref{fig:XRD}(a)) shows only reflections from the epitaxial $\alpha$-Fe$_2$O$_3$ thin film, the Pt layer, and the Al$_2$O$_3$ substrate. No secondary crystalline phases are detected. A broad feature below the $\alpha$-Fe$_2$O$_3$(0006) reflection (grey shaded area) can be assigned to Pt(111) expected at $39.8^\circ$ and points to a textured nature of the Pt top electrode. On an enlarged scale (upper inset in Fig.~\ref{fig:XRD}(a)), satellites due to Laue oscillations are detected around the $\alpha$-Fe$_2$O$_3$(0006) reflection, evidencing a coherent growth with low interface roughness of the $\alpha$-Fe$_2$O$_3$ thin film. The asymmetry on both sides of the $\alpha$-Fe$_2$O$_3$(0006) reflection is caused by interference with the broad Pt(111) reflection. Furthermore, $\alpha$-Fe$_2$O$_3$ shows a low mosaic spread as demonstrated by the full width at half maximum of the rocking curve around the $\alpha$-Fe$_2$O$_3$(0006) reflection of only $0.027^\circ$ (lower inset in Fig.~\ref{fig:XRD}(a)). The in-plane orientation and strain state were investigated by reciprocal space mappings around the $(1\,0\,\overline{1}\,10)$ reflections (Fig.~\ref{fig:XRD}(b)). They reveal the epitaxial relations $[0001]\alpha$-$\mathrm{Fe}_2\mathrm{O}_3 \negthickspace \parallel \negthickspace [0001]\mathrm{Al}_2\mathrm{O}_3$ and $[10\overline{1}0]\alpha$-$\mathrm{Fe}_2\mathrm{O}_3 \negthickspace \parallel \negthickspace [10\overline{1}0]\mathrm{Al}_2\mathrm{O}_3$. We derive lattice constants of $a=0.505$\,nm and $c=1.372$\,nm very close to the respective bulk values, indicating a nearly fully relaxed strain state for our $\alpha$-Fe$_2$O$_3$ films. Furthermore, low interface and surface roughnesses of 0.90\,nm and 0.76\,nm (rms values), respectively, are confirmed by X-ray reflectivity \cite{Supplement}. We note that up to now no clear recipe has been established to prepare a monophase termination of $\alpha$-Fe$_2$O$_3$(0001) \cite{Parkinson:2016}. DFT-based calculations suggest that a Fe termination containing half of the inter-plane Fe is stable at low oxygen pressures \cite{Wang:1998}. Together with an interface roughness of our sample exceeding the interlayer distance of 0.23\,nm, this suggests that Fe$^{3+}$ spins of both magnetic sublattices (i.e.~with opposite directions) are present at the Pt/$\alpha$-Fe$_2$O$_3$ interface. In summary, our $\alpha$-Fe$_2$O$_3$/Pt bilayer is of the same high structural quality as the prototypical ferrimagnetic Y$_3$Fe$_5$O$_{12}$/Pt heterostructures reported earlier \cite{Althammer:2013}.

\section{Angle Dependence of the Spin Hall Magnetoresistance}

\begin{figure}
\includegraphics[width=1.0\columnwidth]{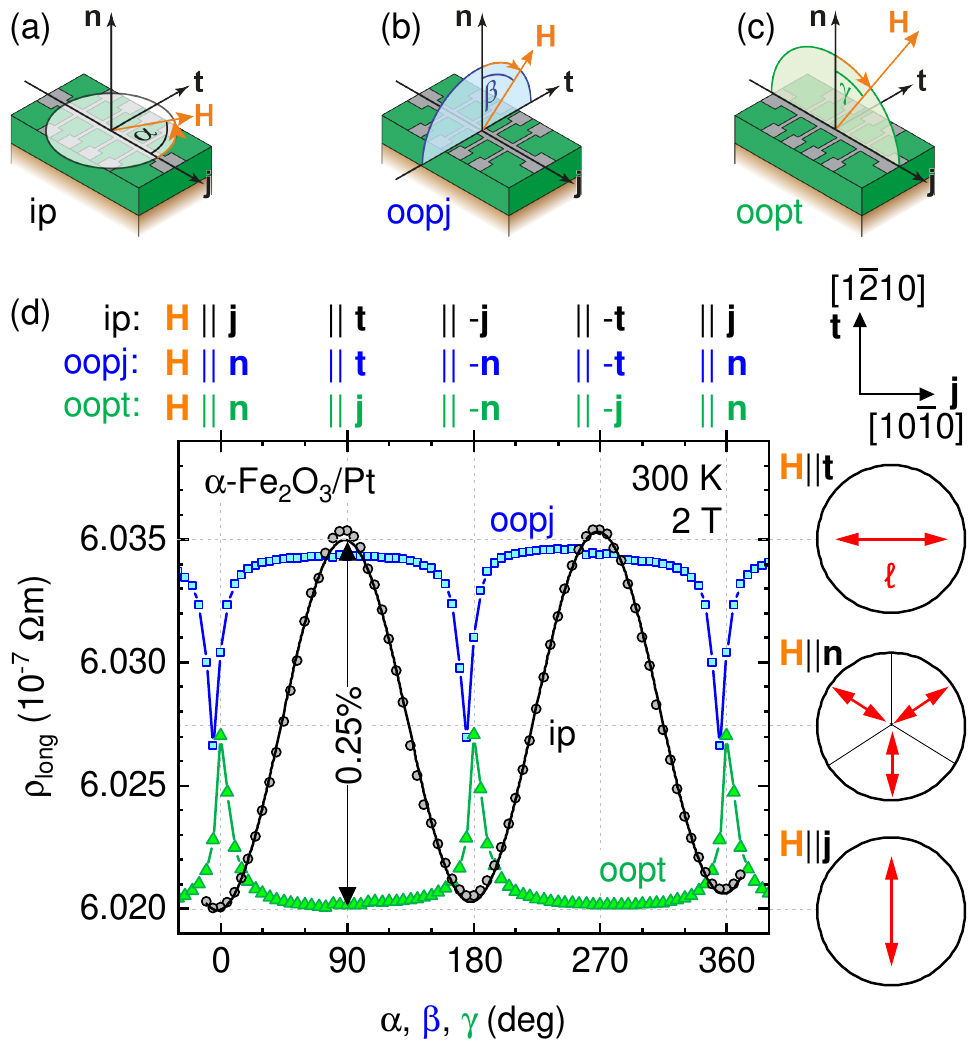}
    \caption{
    ADMR of a (0001)-oriented $\alpha$-Fe$_2$O$_3$/Pt heterostructure. The external magnetic field $\mathbf{H}$ is rotated in three different planes: (a) in the film plane (ip, angle $\alpha$), (b) perpendicular to the current direction $\mathbf{j}$ (oopj, $\beta$), and (c) perpendicular to the transverse direction $\mathbf{t}$ (oopt, $\gamma$). The vector $\mathbf{n}$ denotes the film normal. (d) The longitudinal resistivity $\rho_\mathrm{long}$ is recorded at 300\,K and 2\,T for all three rotation planes: ip (black circles), oopj (blue squares), and oopt (green triangles). The black line is a fit to the ip data according to Eq.~(\ref{eq:rholong-trans1}), the blue and green lines are guides to the eye. The SMR amplitude of $0.25\%$ (vertical black double arrow) is significantly larger than in prototypical Y$_3$Fe$_5$O$_{12}$/Pt structures. At the high (low) resistivity level, $\alpha$-Fe$_2$O$_3$ is in a monodomain state with the Ne\'{e}l vector $\boldsymbol{\ell}$ pointing parallel (perpendicular) to $\mathbf{j}$ with $\mathbf{H}||\mathbf{t}$ ($\mathbf{H}||\mathbf{j}$). For $\mathbf{H}||\mathbf{n}$, only occurring in oopj and oopt geometry, $\alpha$-Fe$_2$O$_3$ exhibits a multidomain state with medium resistivity.  The corresponding domain patterns are illustrated to the right, besides the data plots.
    }
 \label{fig:Chaos}
\end{figure}

In the following, we discuss an $\alpha$-Fe$_2$O$_3$/Pt bilayer sample with thicknesses of $t_\mathrm{Fe_2O_3} = 91.4$\,nm and $t_\mathrm{Pt} = 3.0$\,nm. For transport measurements, a Hall bar-shaped mesa structure with a width of $w\,=\,81\,\mu$m and a longitudinal contact separation (length) of $l\,=\,609\,\mu$m was patterned into the bilayer via photolithography and Ar ion milling (Fig.~\ref{fig:Chaos}). For a dc current of $\pm100\,\mu$A applied in the $[10\overline{1}0]$ direction, the longitudinal ($\rho_\mathrm{long}$) and the transverse ($\rho_\mathrm{trans}$) resistivities are measured in a standard four-probe configuration. A current-reversal method is applied to eliminate thermal effects \cite{Ganzhorn:2016}. We restrict our investigation to room temperature, where the (0001) plane is a magnetic easy plane. We perform angle-dependent magnetoresistance (ADMR) measurements by rotating an external magnetic field of constant magnitude $H$ in three different orthogonal planes of the (0001)-oriented $\alpha$-Fe$_2$O$_3$ using the same notation as in Ref.~\onlinecite{Althammer:2013} and Fig.~\ref{fig:Chaos}(a-c): $(0001) =$ ``ip'' (in-plane, angle $\alpha$, black); $(10\overline{1}0) =$ ``oopj'' (out-of-plane perpendicular to \textbf{j}, angle $\beta$, blue); $(1\overline{2}10) =$ ``oopt'' (out-of-plane perpendicular to \textbf{t}, angle $\gamma$, green).

For ip rotations at $\mu_0H = 2$\,T, $\rho_\mathrm{long}(\alpha)$ displays the characteristic SMR oscillations with $180^\circ$ period (black circles in Fig.~\ref{fig:Chaos}(d)). The minima and maxima are located at $\mathbf{H} \negthickspace \parallel \negthickspace \pm \mathbf{j}$ and $\mathbf{H} \negthickspace \parallel \negthickspace \pm \mathbf{t}$, respectively, representing the signature of the AF (``negative'') SMR \cite{Hoogeboom:2017, Fischer:2018, Baldrati:2018} with a phase shift of $90^\circ$ compared to the ferromagnetic (``positive'') SMR in ferrimagnetic Y$_3$Fe$_5$O$_{12}$/Pt \cite{Althammer:2013, Chen:2013} or $\gamma$-Fe$_2$O$_3$/Pt \cite{Dong:2019}. The SMR amplitude is almost saturated at $\mu_0H = 2$\,T (see below). We can safely assume a monodomain state of the $\alpha$-Fe$_2$O$_3$ thin film with the N\'{e}el vector $\boldsymbol{\ell} = (\mathbf{m}_1 - \mathbf{m}_2)/2$ with $\mathbf{m}_i = \mathbf{M}_i/|\mathbf{M}_i|$
rotating coherently and perpendicular to $\mathbf{H}$ in the magnetic easy (0001) plane. The data are well described by
\begin{align}
  \label{eq:rholong-trans1}
  \rho_\mathrm{long} &= \rho_0 + \frac{\rho_1}{2}(1-\cos2\alpha)
  \\
  \label{eq:rholong-trans2}
  \rho_\mathrm{trans} &= - \frac{\rho_3}{2}\sin2\alpha
\end{align}
(black line in Fig.~\ref{fig:Chaos}(d)) with $\rho_0$ approximately equal to the normal resistivity of the Pt layer and $\rho_1$ and $\rho_3$ representing the longitudinal and the transverse SMR coefficients, respectively \cite{Supplement}, as demonstrated earlier for AF NiO/Pt \cite{Fischer:2018}. However, the SMR amplitude of $0.25\%$ for $\alpha$-Fe$_2$O$_3$/Pt is more than a factor of 3 higher than for NiO/Pt and, remarkably, even twice as large as for the prototypical ferrimagnetic Y$_3$Fe$_5$O$_{12}$/Pt heterostructures with similar Pt thickness \cite{Althammer:2013}. In fact, it is larger than for any other reported bilayer compound so far. We attribute this to the large density of magnetic Fe$^{3+}$ ions in $\alpha$-Fe$_2$O$_3$. With a spin Hall angle of 0.11 and a spin diffusion length of 1.5\,nm for Pt \cite{Meyer:2014}, we calculate $G_r \! = \! 1.38 \! \times \! 10^{15} \, \mathrm{\Omega}^{-1} \mathrm{m}^{-2}$ for the real part of the spin mixing interface conductance. Although the situation regarding the magnitude of $G_r$ is confusing since the values reported in the literature are obtained from different techniques and are not fully comparable to each other \cite{Zhu:2019}, our value is of the order of all-metallic ferromagnetic interfaces \cite{Yang:2016, Czeschka:2011}. It is consistent with that reported by Cheng and coworkers for antiferromagnetic $\alpha$-Fe$_2$O$_3$/Pt \cite{Cheng:2019}, and about one order of magnitude larger than for ferrimagnetic Y$_3$Fe$_5$O$_{12}$/Pt heterostructures \cite{Althammer:2013, Meyer:2014}.

The ADMR of the out-of-plane rotations is qualitatively different. For oopj rotations of $\mathbf{H}$, we observe $\rho_\mathrm{long}$ in the maximum resistive state for a wide range of angles $\beta$ around $\mathbf{H} \negthickspace \parallel \negthickspace \pm \mathbf{t}$ (blue squares in Fig.~\ref{fig:Chaos}(d)), indicating a monodomain state with $\boldsymbol{\ell}\negthickspace \parallel \negthickspace \pm \mathbf{j}$. For the oopt geometry, on the other hand, $\rho_\mathrm{long}$ stays in the minimum resistive state for a wide range of angles $\gamma$ around $\mathbf{H} \negthickspace \parallel \negthickspace \pm \mathbf{j}$ (green triangles in Fig.~\ref{fig:Chaos}(d)), indicating again a monodomain state with $\boldsymbol{\ell}\negthickspace \parallel \negthickspace \pm \mathbf{t}$. Both observations show that $\boldsymbol{\ell}$ does not follow $\mathbf{H}$ for out-of-plane rotations. $\rho_\mathrm{long}$ changes significantly only close to $\mathbf{H} \negthickspace \parallel \negthickspace \pm \mathbf{n}$: both oopj and oopt curves meet for magnetic fields perpendicular to the sample surface at the midpoint of the two states with extremal resistance. According to the SMR model for a multidomain antiferromagnet \cite{Fischer:2018}, we interpret this observation with the ``decay'' of a monodomain into a three-domain state when $\mathbf{H}$ points orthogonal to the magnetic easy (0001) plane of $\alpha$-Fe$_2$O$_3$, in agreement with a recent preprint \cite{Cheng:2019}.

\section{Field Dependence of the Spin Hall Magnetoresistance}

\begin{figure}
\includegraphics[width=1.0\columnwidth]{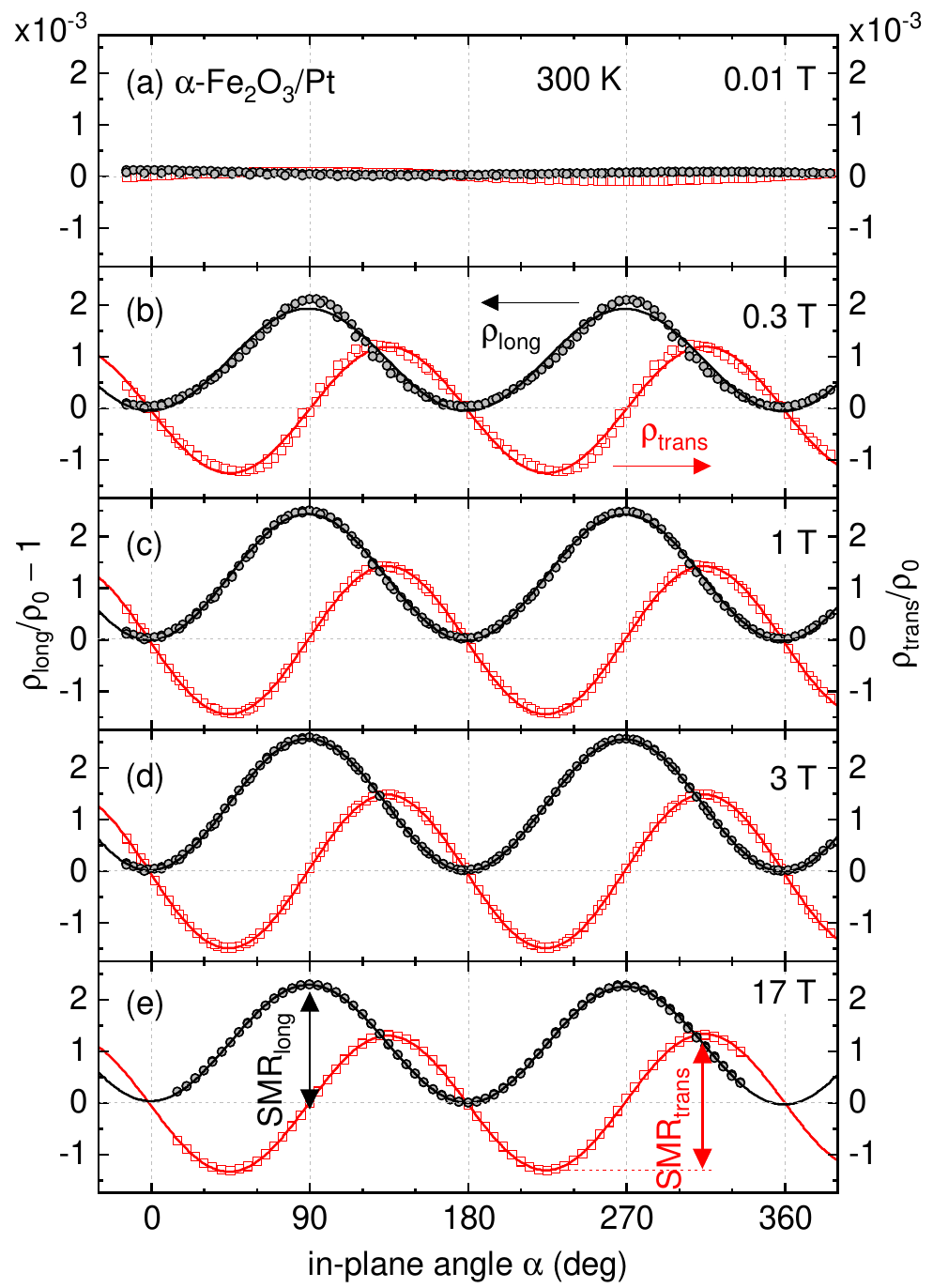}
    \caption{
    In-plane ADMR of a (0001)-oriented $\alpha$-Fe$_2$O$_3$/Pt heterostructure at 300\,K in magnetic fields of (a) 0.01\,T, (b) 0.3\,T, (c) 1\,T, (d) 3\,T, and (e) 17\,T. The normalized longitudinal ($\rho_\mathrm{long}$, black circles, left axis) and transverse resistivities ($\rho_\mathrm{trans}$, red squares, right axis) are plotted as a function of the in-plane magnetic field orientation $\alpha$. The lines are fits to the data using $\cos2\alpha$ and $\sin2\alpha$ functions analogous to Eqs.~(\ref{eq:rholong-trans1} \& (\ref{eq:rholong-trans2})).
    }
 \label{fig:rho-long-trans}
\end{figure}

To obtain further insight into the AF domain configurations, we perform ip ADMR measurements of both $\rho_\mathrm{long}$ and $\rho_\mathrm{trans}$ at different magnitudes of the magnetic field from 10\,mT to 17\,T (Fig.~\ref{fig:rho-long-trans}). The expected ($-\cos2\alpha$) and ($-\sin2\alpha$) dependencies of $\rho_\mathrm{long}$ and $\rho_\mathrm{trans}$, respectively, are clearly observed for $\mu_0H \geq 300$\,mT (Fig.~\ref{fig:rho-long-trans}(b-e)). This angular dependence is fully consistent with the model introduced earlier for NiO/Pt \cite{Fischer:2018} and Eqs.~(\ref{eq:rholong-trans1}) \& (\ref{eq:rholong-trans2}) and clearly shows that our $\alpha$-Fe$_2$O$_3$ is AF with the resistivity of Pt being sensitive to $\boldsymbol{\ell}$, which rotates coherently in the easy (0001) plane perpendicular to $\mathbf{H}$. The data is further fully consistent with recent experiments in Pt on canted ferrimagnets, where the same angular dependence is observed close to the compensation temperature \cite{Ganzhorn:2016}, as well as experiments in Y$_3$Fe$_5$O$_{12}$/NiO/Pt \cite{Shang:2016,Hou:2017, Lin:2017, Hung:2017} and NiO/Pt \cite{Hoogeboom:2017, Fischer:2018, Baldrati:2018}. For $\mu_0H \lesssim 100$\,mT, the applied field is smaller than $\mu_0H_\mathrm{MD}$, resulting in hardly detectable SMR oscillations (Fig.~\ref{fig:rho-long-trans}(a)).

\begin{figure}
\includegraphics[width=0.9\columnwidth]{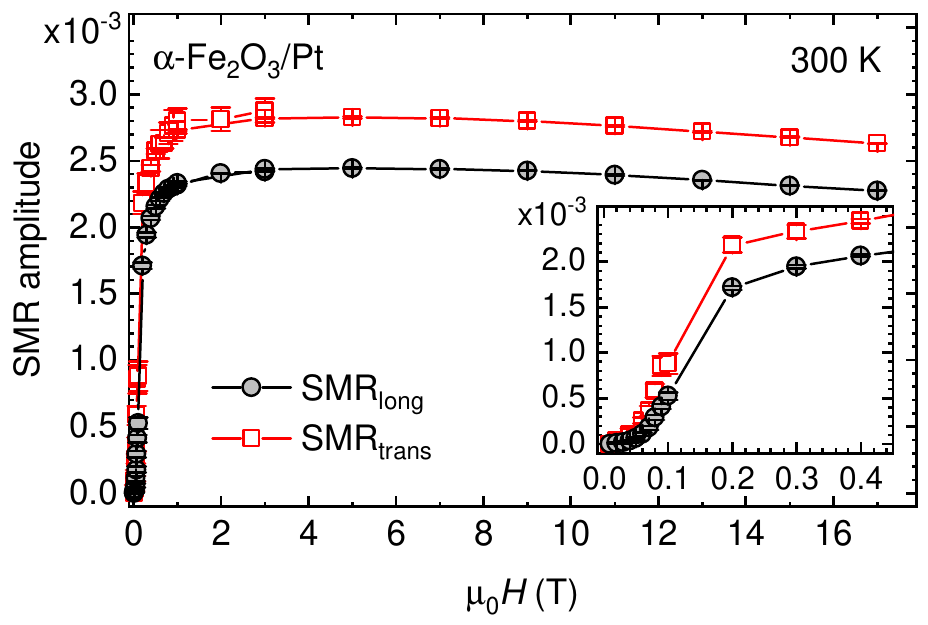}
    \caption{
    SMR amplitudes $\mathrm{SMR_{long}}$ (black circles) and $\mathrm{SMR_{trans}}$ (red squares) of a (0001)-oriented $\alpha$-Fe$_2$O$_3$/Pt heterostructure derived from ADMR measurements at 300\,K in different external magnetic fields $H$ (cf.~Fig.~\ref{fig:rho-long-trans}).
    }
 \label{fig:amplitude}
\end{figure}

For a detailed analysis of the field dependence of $\rho_\mathrm{long}$ and $\rho_\mathrm{trans}$, we fit our data analogous to Eqs.~(\ref{eq:rholong-trans1}) \& (\ref{eq:rholong-trans2}) using $\cos2\alpha$ and $\sin2\alpha$ functions, respectively, (solid lines in Fig.~\ref{fig:rho-long-trans}) and plot the SMR amplitudes $\mathrm{SMR_{long}}$ and $\mathrm{SMR_{trans}}$ (double arrows in Fig.~\ref{fig:rho-long-trans}(e)) as a function of the magnetic field magnitude in Fig.~\ref{fig:amplitude}. Remarkably, $\mathrm{SMR_{trans}}$ exceeds $\mathrm{SMR_{long}}$ for fields above 100\,mT. This unexpected observation may indicate the presence of large $180^\circ$ domains in $\alpha$-Fe$_2$O$_3$, exceeding the width of the Hall bar.
Thick
$180^\circ$ domain walls, present only along the length of the Hall bar, might then effectively reduce the SMR in the longitudinal voltage, but not in the transverse one.
We note that such large domains and thick domain walls have been reported for (0001)-oriented bulk material at moderate magnetic fields \cite{Eaton:1969, Morrish:1994, Clark:1983}, but cannot be resolved in (0001)-oriented thin films at room temperature in zero magnetic field \cite{Ross:2019}.

Furthermore, the field evolution of the SMR amplitude is qualitatively different from the one in AF NiO/Pt \cite{Fischer:2018}. In $\alpha$-Fe$_2$O$_3$/Pt, we find that both $\mathrm{SMR_{long}}$ and $\mathrm{SMR_{trans}}$ saturate already around 3\,T and then gradually decrease again from 5\,T to 17\,T. This gradual decrease can be traced back to an increasing canting of the AF sublattices thus reducing the value of $\boldsymbol{\ell}$ and an emerging non-zero net $\mathbf{M}$ \cite{Supplement}. The fast saturation, on the other hand, points to a lower destressing energy compared to NiO. The field dependence of the SMR amplitude indicates that the $120^\circ$ domains in our $\alpha$-Fe$_2$O$_3$ thin film vanish at $\simeq 3$\,T where the SMR amplitude starts to saturate (Fig.~\ref{fig:amplitude}). To quantify the destressing effects, we identify 3\,T with the mono-domainization field $H_\mathrm{MD}$, since the leftover $180^\circ$ domains have indistinguishable destressing energy density. With an exchange field of $\mu_0H_\mathrm{ex} = 900$\,T \cite{Morrish:1994, Bodker:2000}, we derive a destressing field $\mu_0H_\mathrm{dest} = \mu_0 H^2_\mathrm{MD} / (4H_\mathrm{ex}) \simeq 2.5$\,mT, smaller than the 46\,mT in epitaxial NiO thin films on Al$_2$O$_3$ \cite{Fischer:2018}. This value is reasonable, since the magnetostriction $\lambda = 4\times10^{-6}$ in the basal plane of $\alpha$-Fe$_2$O$_3$ at 293\,K \cite{Voskanyan:1968} is by a factor of $\sim20$ smaller than $\lambda = (9\pm3)\times10^{-5}$ in NiO \cite{Yamada:1966b}.

\section{Conclusion and Outlook}

In summary, we present a detailed investigation of the SMR in antiferromagnetic $\alpha$-Fe$_2$O$_3$/Pt heterostructures at room temperature studying three orthogonal rotation planes of the magnetic field. We consistently describe the angular dependence of the data in a three-domain model, considering a field-dependent canting of the AF sublattices. Our data supports the picture that each magnetic sublattice contributes separately to the SMR. Surprisingly, we find a large SMR amplitude of $0.25\%$. This value well exceeds the established values for Y$_3$Fe$_5$O$_{12}$/Pt or any other Pt-based thin film heterostructures reported in literature so far. AF materials are therefore expected to play an important role in SMR-related research and applications. Due to the small destressing field, the SMR amplitude reaches $0.20\%$ (corresponding to $^4\!/_5$ of its maximum value) already at 300\,mT, i.e.~at much smaller magnetic fields than in comparable AF NiO/Pt heterostructures \cite{Fischer:2018}. This combination of high sensitivity at low magnetic fields and room temperature operation makes $\alpha$-Fe$_2$O$_3$/Pt a promising material system both for a viable SMR source and future spin transfer torque based devices. The large spin mixing interface conductance of $1.38 \! \times \! 10^{15} \, \mathrm{\Omega}^{-1} \mathrm{m}^{-2}$ makes it further suitable for spin current-induced magnetization switching or other spin transfer torque-based applications in the emerging field of antiferromagnetic spintronics.

\section*{Acknowledgments}

We thank T.~Brenninger, A.~Habel, and K.~Helm-Knapp for technical support as well as O.~Gomonay and A.~Kamra for fruitful discussions. We gratefully acknowledge financial support of the German Research Foundation via Germany's  Excellence Strategy (EXC-2111-390814868). N.V. acknowledges support from a Laura-Bassi stipend of the Technical University of Munich.

\clearpage

\section*{Supplemental Material}

For completion and confirmation of our results, we present here additional data from the $\alpha$-Fe$_2$O$_3$/Pt heterostructure investigated in the main text (in the following referred to as ``sample\#1'') with layer thicknesses of $t_\mathrm{Fe_2O_3} = 91.4$\,nm and $t_\mathrm{Pt} = 3.0$\,nm for $\alpha$-Fe$_2$O$_3$ and Pt, respectively, as well as additional data from a second sample with $t_\mathrm{Fe_2O_3} = 108.5$\,nm without Pt top layer (``sample\#2'') and a third sample with $t_\mathrm{Fe_2O_3} = 61.5$\,nm and $t_\mathrm{Pt} = 1.8$\,nm (``sample\#3''). All samples are fabricated as described in the main text on (0001)-oriented Al$_2$O$_3$ substrates. We further summarize the mathematical description of the in-plane SMR oscillations in systems with two antiparallel sublattice magnetizations.

\section*{Supplemental Material: Interface and Surface Characterization}

\begin{figure}[b]
\includegraphics[width=0.9\columnwidth]{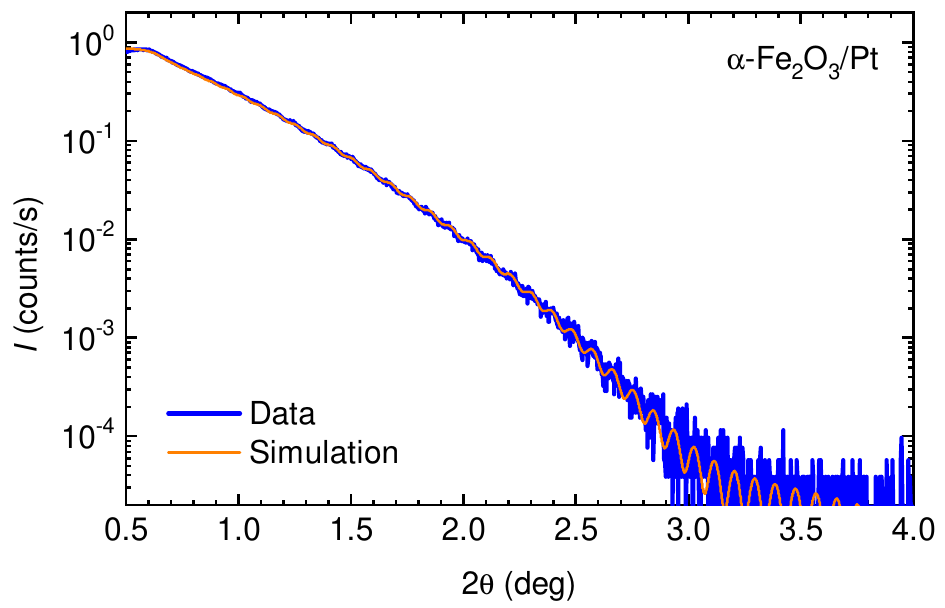}
    \caption{
    X-ray reflectivity data (blue line) from the $\alpha$-Fe$_2$O$_3$/Pt heterostructure investigated in the main text, fabricated on a (0001)-oriented Al$_2$O$_3$ substrate (sample\#1). The red line shows the simulation of the data assuming thicknesses of 91.4\,nm and 3.0\,nm for the $\alpha$-Fe$_2$O$_3$ and Pt layers, respectively.
 }
\label{fig:XRR}
\end{figure}

We determine the thicknesses and roughnesses of our bilayer samples via high-resolution X-ray reflectivity. A simulation of the data from sample\#1 (Fig.~\ref{fig:XRR}) reveals 91.4\,nm and 3.0\,nm for the thicknesses of the $\alpha$-Fe$_2$O$_3$ and the Pt layer, respectively. The interface roughness is found to be 0.90\,nm (rms value) and the surface roughness 0.76\,nm (rms value).

\begin{figure}
\includegraphics[width=0.75\columnwidth]{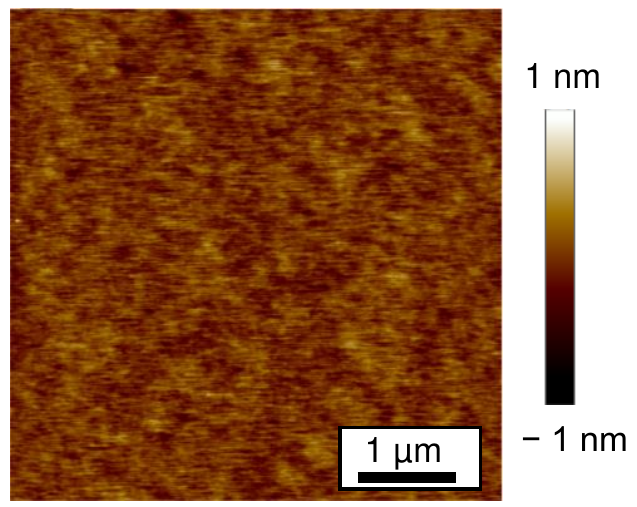}
    \caption{
    Atomic force microscopy (AFM) image of a single $\alpha$-Fe$_2$O$_3$ thin film with a thickness of 108.5\,nm, fabricated on a (0001)-oriented Al$_2$O$_3$ substrate (sample\#2).
 }
\label{fig:AFM}
\end{figure}

Additionally, we investigate the surface morphology on the micrometer scale of a second sample without Pt top layer (sample\#2). The AFM image shows a smooth surface over $5 \times 5\,\mu$m$^2$ (Fig.~\ref{fig:AFM}). A careful analysis reveals an $\alpha$-Fe$_2$O$_3$ surface roughness of 0.13\,nm (rms value).

\section*{Supplemental Material: Magnetic Characterization}

\begin{figure}
\includegraphics[width=0.9\columnwidth]{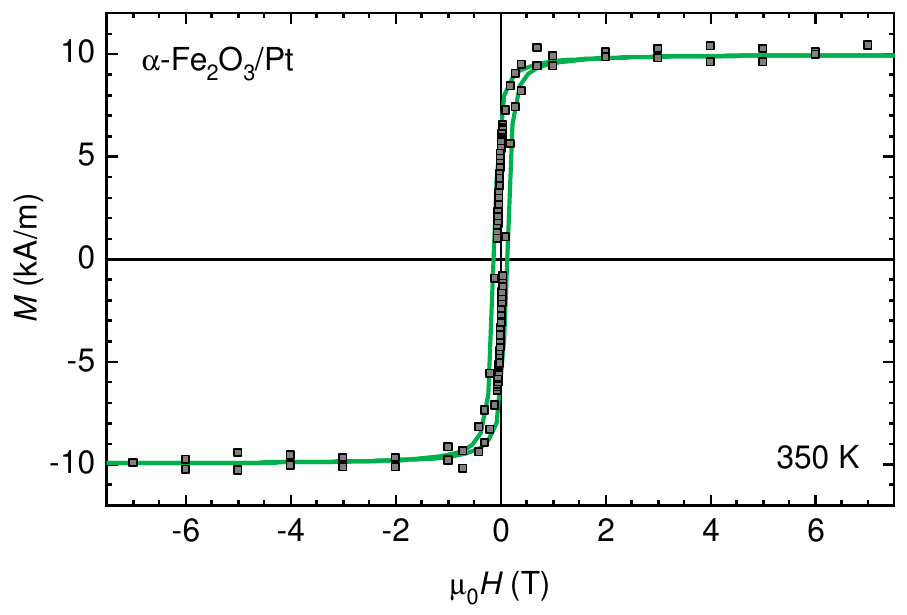}
    \caption{
    Magnetic hysteresis of sample\#1. The magnetization $M$ (black squares) is plotted as a function of the magnetic field $H$, applied in the film plane, at 350\,K. A linear background, mainly originating from the diamagnetic substrate, was subtracted from the data. The green lines are guides to the eye.
 }
\label{fig:SQUID}
\end{figure}

We measure the magnetization $M$ of the $\alpha$-Fe$_2$O$_3$/Pt heterostructures via superconducting quantum interference device (SQUID) magnetometry at $T = 350$\,K. The magnetic field of up to $\mu_0 H = 7$\,T is applied in the film plane, i.e.~in the magnetically easy (0001) plane of $\alpha$-Fe$_2$O$_3$. After subtracting a linear background, originating mainly from the diamagnetic Al$_2$O$_3$ substrate, we observe a step-like behavior in the $M$ versus $H$ curve with a small hysteresis around zero field and a low saturation magnetization of $M_\mathrm{s} \simeq 10$\,kA/m (Fig.~\ref{fig:SQUID}). This value is compatible with the canted arrangement of the antiferromagnetically ordered sublattices in the (0001) planes at $T > T_\mathrm{M}$. From $S=5/2$ for Fe$^{3+}$ and with an ion density of $n_\mathrm{Fe^{3+}} = 39.81$\,nm$^{-3}$ \cite{Springer-Materials:Fe2O3}, we deduce a canting angle of the magnetic sublattices of $0.31^\circ$ away from their antiferromagnetic orientation in agreement with values reported for $\alpha$-Fe$_2$O$_3$ bulk samples \cite{Morrish:1994} and nanoparticles \cite{Bodker:2000}.

We note that the diamagnetic signal from the substrate is not known accurately enough to allow for an exact subtraction. We instead remove a linear background from the magnetization data which contains also other contributions, i.e.~the Pauli paramagnetism of Pt and the field-dependent canting effect in $\alpha$-Fe$_2$O$_3$. Therefore, the magnetization data in Figure~\ref{fig:SQUID} cannot fully reproduce the field dependence of the SMR amplitude (cf.~Fig.~4 of the main Letter), saturates earlier at $\mu_0 H_\mathrm{s} = 700$\,mT, and does not increase at large magnetic fields.

\section*{Supplemental Material: Hall Bar Geometry}

\begin{figure}
\includegraphics[width=0.9\columnwidth]{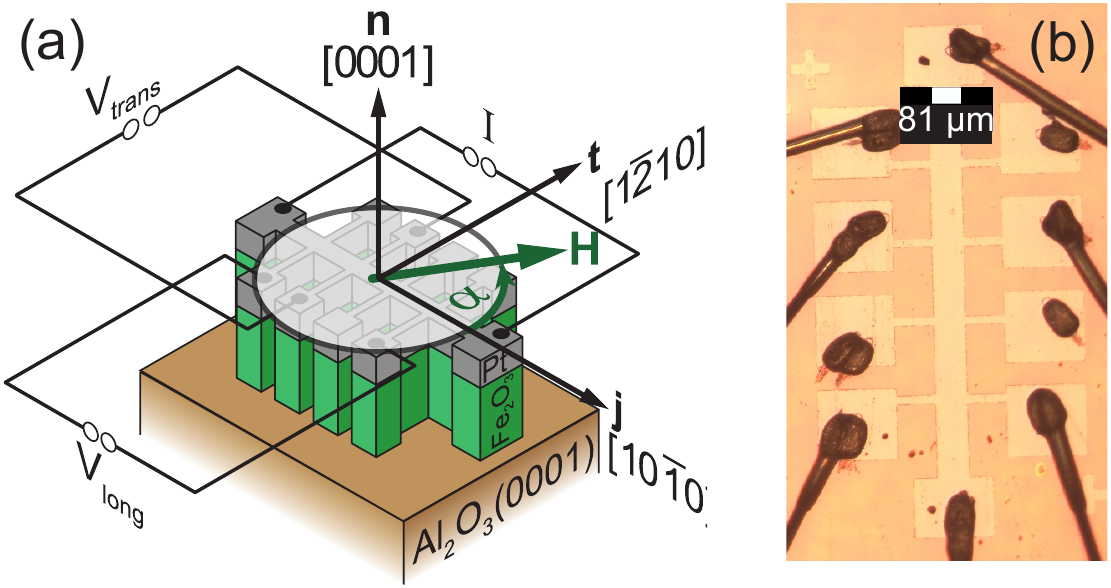}
    \caption{
    (a) Schematic view of the Hall bar mesa structure. For details see text.
    (b) Micrograph image (top view) of the patterned Hall bar on sample\#1, showing the geometrical dimensions and 10 bond pads with bonding wires.
 }
\label{fig:Hallbar}
\end{figure}

For the transport measurements, we use a Hall bar geometry to apply an electrical current $I$ and determine the longitudinal and transverse voltages $V_\mathrm{long}$ and $V_\mathrm{trans}$ in a standard four-probe configuration (Fig.~\ref{fig:Hallbar}(a)). From the measured $V_\mathrm{long}$ and $V_\mathrm{trans}$, we calculate the longitudinal and transverse resistivities $\rho_\mathrm{long}$ and $\rho_\mathrm{trans}$ according to
\begin{align}
    \rho_\mathrm{long} &= \frac{V_\mathrm{long}wt}{Il}
    \\
    \rho_\mathrm{trans} &= \frac{V_\mathrm{trans}t}{I}
\end{align}
where $l$, $w$, and $t$ are the length, the width, and the thickness of the metallic Pt layer of the Hall bar, respectively. The Hall bar is patterned via photolithography and Ar ion milling with the nominal dimensions $l=600\,\mu$m and $w=80\,\mu$m, resulting in an aspect ratio of $l/w=7.5$. The real dimensions were determined afterwards from an optical micrograph image (Fig.~\ref{fig:Hallbar}(b)) to $l=609\,\mu$m and $w=81\,\mu$m, maintaining the same aspect ratio.

\section*{Supplemental Material: Mathematical Description of the in-plane SMR Oscillations}

In bilayer heterostructures consisting of a heavy metal (e.g.~Pt) on a ferrimagnetic (e.g.~Y$_3$Fe$_5$O$_{12}$) or antiferromagnetic (e.g.~$\alpha$-Fe$_2$O$_3$) insulator with 2 magnetic sublattices, the modulation of the resistivity tensor $\boldsymbol{\rho}$ of the metallic layer due to the SMR effect depends on the directions $\mathbf{m}^{(1)}$ and $\mathbf{m}^{(2)}$ of the magnetizations of each magnetic sublattice \cite{Ganzhorn:2016}. The diagonal component of $\boldsymbol{\rho}$ along the charge current direction $\mathbf{j}$, i.e.~the longitudinal resistivity $\rho_{\mathrm{long}}$, is then given by \cite{Ganzhorn:2016, Chen:2013}
\begin{align}
\rho_{\mathrm{long}}&= \rho_{0}+\frac{1}{2}\sum_{i=1}^2 \rho_1^{(i)} \left[ 1 - \left(\mathbf{m}^{(i)}) \cdot \mathbf{t}\right)^{2}\right] \nonumber \\
	&= \rho_{0}+\frac{1}{2}\sum_{i=1}^2 \rho_1^{(i)} \left[1 - \left(m_t^{(i)}\right)^2 \right]
	\; ,
\end{align}
where $\rho_{0}$ is approximately equal to the normal resistivity of the metallic layer \cite{Chen:2013} and $\rho_1^{(i)}$ represent the SMR coefficients of the magnetic sublattices with $\rho_1^{(i)} \ll \rho_0$. $m_t^{(i)}$ denote the projections of $\mathbf{m}^{(i)}$ on the transverse direction $\mathbf{t}$ (perpendicular to $\mathbf{j}$ in the $\mathbf{j}$-$\mathbf{t}$-interface plane, see Fig.~\ref{fig:Hallbar}(a)).

From a similar consideration, the transverse resistivity $\rho_\mathrm{trans}$ is given by \cite{Chen:2013, Ganzhorn:2016, Althammer:2013}
\begin{align}
  \rho_{\mathrm{trans}}	= \frac{1}{2}\sum_{i=1}^2 \rho_3^{(i)} \, m_j^{(i)} \, m_t^{(i)}
\end{align}
with the transverse SMR coefficients $\rho_3^{(i)} \ll \rho_0$. $m_j^{(i)}$ are the projections of $\mathbf{m}^{(i)}$ on the current direction $\mathbf{j}$.

When defining $\varphi^{(i)}$ as the angle between $\mathbf{j}$ and $\mathbf{m}^{(i)}$ and assuming that the sublattice magnetizations stay in the $\mathbf{j}$-$\mathbf{t}$ plane, $\rho_{\mathrm{long}}$ and $\rho_{\mathrm{trans}}$ depend on $\varphi^{(i)}$ as
\begin{align}
  \rho_{\mathrm{long}}  &= \rho_{0} +  \frac{1}{2} \sum_{i=1}^2 \rho_1^{(i)} \left[1 + \cos2\varphi^{(i)}\right]
  \label{eq:rho_long}
  \\
  \rho_{\mathrm{trans}}	&=             \frac{1}{2} \sum_{i=1}^2 \rho_3^{(i)}           \sin2\varphi^{(i)}
  \label{eq:rho_trans}
	\; .
\end{align}
In the above description, any canting between the magnetic sublattices is neglected such that they are oriented antiparallel with $\varphi^{(2)} = 180^\circ+\varphi^{(1)}$.

We now apply a rotating external magnetic field $\mathbf{H}$ in the $\mathbf{j}$-$\mathbf{t}$ plane where $\alpha$ is the angle between $\mathbf{j}$ and $\mathbf{H}$ (see Fig.~\ref{fig:Hallbar}(a)). Neglecting any magnetic anisotropy or domain effects, the net magnetization and thus the sublattice magnetizations in ferrimagnets will follow $\mathbf{H}$ and $\varphi^{(1)} \equiv \alpha$ \cite{Althammer:2013}. Eqs.~(\ref{eq:rho_long}) and (\ref{eq:rho_trans}) then read
\begin{align}
  \rho_{\mathrm{long}}  &= \rho_{0} + \frac{\rho_1}{2} \left[1 + \cos2\alpha\right]
  \\
  \rho_{\mathrm{trans}}	&=            \frac{\rho_3}{2}           \sin2\alpha
\end{align}
with $\rho_1 = \rho_1^{(1)}+\rho_1^{(2)}$ and $\rho_3 = \rho_3^{(1)}+\rho_3^{(2)}$.

In antiferromagnets, however, the sublattices are oriented perpendicular to $\mathbf{H}$, resulting in $\varphi^{(1)} \equiv 90^\circ + \alpha$ \cite{Fischer:2018}. Eqs.~(\ref{eq:rho_long}) and (\ref{eq:rho_trans}) then read \cite{Fischer:2018}
\begin{align}
  \rho_{\mathrm{long}}  &= \rho_{0} + \frac{\rho_1}{2} \left[1 - \cos2\alpha\right]
  \\
  \rho_{\mathrm{trans}}	&=          - \frac{\rho_3}{2}           \sin2\alpha
	\; .
\end{align}
Because of the minus signs in these equations, the SMR in antiferromagnets is sometimes referred to as \textit{negative} spin Hall magnetoresistance.

From fits of the experimental data to the above equations, we finally determine the SMR amplitudes
\begin{align}
  \mathrm{SMR_{long}}  &= \frac{\rho_{1}}{\rho_0}
  \\
  \mathrm{SMR_{trans}} &= \frac{\rho_{3}}{\rho_0}
\end{align}
which we analyze as a function of the magnitude $H$ of the external magnetic field.

\section*{Supplemental Material: Longitudinal and Transverse SMR Amplitudes}

\begin{figure}[b]
\includegraphics[width=0.9\columnwidth]{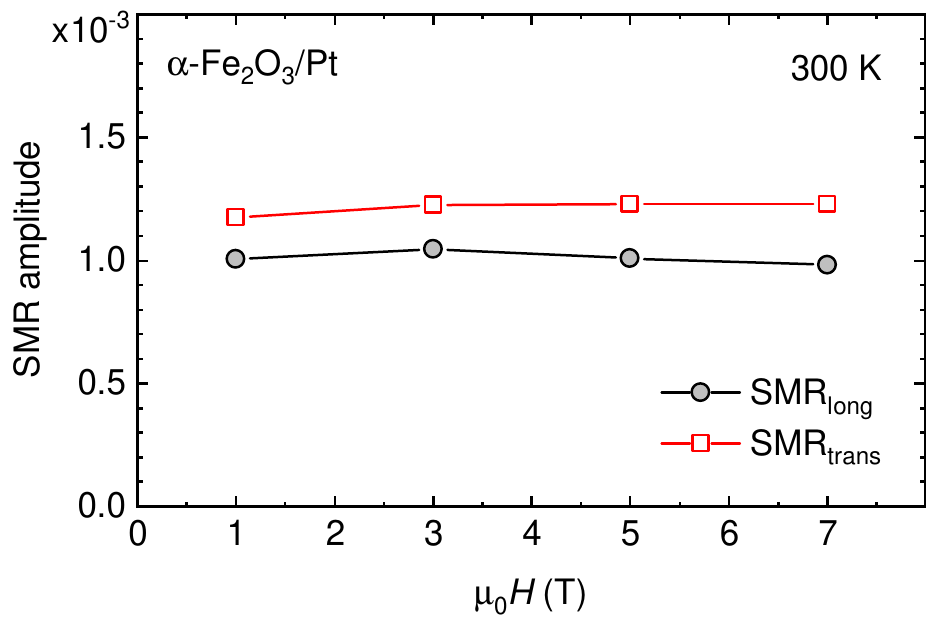}
    \caption{
    SMR amplitudes $\mathrm{SMR_{long}}$ (black circles) and $\mathrm{SMR_{trans}}$ (red squares) of sample\#3, derived from in-plane ADMR measurements at room temperature in different external magnetic fields $H$.
 }
\label{fig:SMR}
\end{figure}

The observed discrepancy between the longitudinal and the transverse SMR amplitudes $\mathrm{SMR_{long}}$ and $\mathrm{SMR_{trans}}$ is a robust feature in our samples. To address this behavior, we investigate another (0001)-oriented $\alpha$-Fe$_2$O$_3$/Pt heterostructure (sample\#3) with the same dimensions of the Hall bar, but with a thinner Pt layer ($t_\mathrm{Pt} = 1.8$\,nm). Again, we apply a dc current of $100\,\mu$A while rotating a magnetic field of up to 7\,T in the film plane at 300\,K. Compared to sample\#1, the overall SMR signal is only about half as large (Fig.~\ref{fig:SMR}). This becomes clear when recalling that the SMR amplitude crucially depends on the thickness of the Pt layer and drastically decreases when it becomes thinner than twice of the spin diffusion length in Pt, as reported earlier for Y$_3$Fe$_5$O$_{12}$/Pt \cite{Althammer:2013, Meyer:2014}. Even more important, however, the transverse SMR amplitude again exceeds the longitudinal one ($\mathrm{SMR_{long}}$) and both saturate around 3\,T (Fig.~\ref{fig:SMR}), like in sample\#1. This behavior fully reproduces the previous results described in the main Letter. In a smaller Hall bar with $160\,\times\,2\,\mu\mathrm{m}^2$ dimensions (not shown here), however, we did not observe this discrepancy. Moreover, we note that we used the same $600\,\times\,80\,\mu\mathrm{m}^2$ Hall bar lithography mask and patterning process as well as the same deposition chambers and magnetotransport setups in our earlier SMR studies of Y$_3$Fe$_5$O$_{12}$/Pt \cite{Althammer:2013} and NiO/Pt \cite{Fischer:2018} and have never observed such a discrepancy before. In view of this, our results from sample\#3 are fully consistent and confirm our assumption of the main Letter that $180^\circ$ domains larger than the width of the Hall bar but smaller than its length are present in our $\alpha$-Fe$_2$O$_3$ thin films.

\bibliography{SMR_Fe2O3_Pt}

\end{document}